\documentclass[aps,prb,reprint,amsmath,amssymb,floatfix,superscriptaddress]{revtex4-2}

\usepackage{float}
\usepackage{graphicx}
\usepackage{physics}
\usepackage{color}
\usepackage{times,psfrag}
\usepackage{dsfont}
\usepackage{dcolumn}
\usepackage{bm,bbm}
\usepackage{latexsym,amsmath,amssymb, bm,euscript}
\usepackage[normalem]{ulem}
\usepackage[hidelinks]{hyperref}
\usepackage{microtype}
\usepackage{verbatim} 

\newcommand{\SiO}{SiO$_{2}$}

\newcommand{\zbAgI}{$\gamma$-AgI}
\newcommand{\wzAgI}{$\beta$-AgI}

\begin{document}

\title{Nonlinear Anisotropy in Phase-Tuned Wide-Gap Halides}

\author{L. Landivar Scott}
\affiliation{Department of Physics, University of Arkansas, AR 72701, USA}
\author{L. M. Vogl}
\affiliation{Max Planck Institute for Sustainable Materials, 40237 D{\"u}sseldorf, Germany}
\author{C. Klenke}
\affiliation{Department of Materials Science and Engineering, University of Arkansas, AR 72701, USA}
\author{S. Puri}
\affiliation{Department of Physics, University of Arkansas, AR 72701, USA}
\author{H. Nakamura}
\email{hnakamur@uark.edu}
\affiliation{Department of Physics, University of Arkansas, AR 72701, USA}
\affiliation{Department of Materials Science and Engineering, University of Arkansas, AR 72701, USA}

\begin{abstract}
Silver iodide (AgI) thin films offer a compelling platform for studying nonlinear optical phenomena due to their intrinsic noncentrosymmetric lattice and direct band gap. Here, we investigate the nonlinear optical properties of AgI thin films grown by physical vapor deposition that selectively produce zincblende (\zbAgI) and wurtzite (\wzAgI) phases. Using a combination of polarization-resolved second harmonic generation (SHG) and two-photon photoluminescence (2PPL) spectroscopy, we identify clear phase- and morphology-dependent anisotropic nonlinear responses. Triangular \zbAgI $(111)$ flakes exhibit sixfold SHG symmetry and isotropic 2PPL emission, while rod-shaped \wzAgI $(101)$ samples display twofold-symmetric patterns in both SHG and 2PPL, which are explained by theories that integrate explicit polarization analysis using second- and third-order nonlinear susceptibilities. We estimate $\chi^{(2)}_\mathrm{eff}$ of 0.45 pm/V (\zbAgI) and 0.16 pm/V (\wzAgI), respectively, after correcting for multiple internal reflections and generation of SHG within the AgI film and optical interference effect in the dielectric layers. These results identify AgI as a useful single-composition halide platform for studying phase-dependent nonlinearity and establish a realistic methodology for evaluating nonlinear susceptibilities in layered materials or thin films supported by a substrate.
\end{abstract}

\keywords{Silver Iodide, AgI, Metal Halides, Two Photon Photoluminescence, 2PPL, Second Harmonic Generation, SHG}

\maketitle 

\section{Introduction\label{intro}}

Nonlinear optical effects in crystalline materials provide powerful means of probing symmetry, band structure, and light–matter interactions. In particular, second harmonic generation (SHG) and two-photon photoluminescence (2PPL) have become indispensable tools for studying noncentrosymmetric materials and their symmetry-dependent responses \cite{Boyd2020Book}. The past decade has witnessed remarkable advances in two-dimensional (2D) materials such as transition-metal dichalcogenides (TMDs) and hexagonal boron nitride, where interlayer stacking and rotational alignment can reversibly break or restore inversion symmetry, thereby modulating the nonlinear optical response \cite{mak2010atomically,li2013probing,seyler2015electrical}. Such stacking-dependent SHG effects demonstrate that structural symmetry control at the atomic scale provides a robust handle for engineering nonlinearities in low-dimensional systems.

Beyond van der Waals semiconductors, halide compounds are emerging as a promising new class of nonlinear materials \cite{saouma2017selective,yumoto2024electrically,xu2020halide}. Their large polarizability, tunable electronic structure, and propensity to form noncentrosymmetric lattices give rise to strong  $\chi^{(2)}$ and $\chi^{(3)}$ responses even in bulk phases. Lead halide perovskites and copper or silver halides have exhibited tunable second- and third-order nonlinearities due to structural phase transitions, excitonic resonances, and strain  \cite{goldmann1977band, dinges1976two, alexander2021anisotropic, nakamura2024strongly}. However, in many halide systems, structural changes are accompanied by variations in chemistry, defect populations, or disorder, which complicates attributing the nonlinear optical response solely to the crystallographic phase. A material in which phase can be varied while composition remains fixed is therefore especially useful for separating structure-driven changes in nonlinear anisotropy from compositional effects. Silver iodide (AgI) provides such a platform. AgI exhibits polymorphism between zincblende (\zbAgI) and wurtzite (\wzAgI) under ambient conditions \cite{burley1963polymorphism,patnaik1998studies,sabath2022atomic}, while preserving the same Ag-I stoichiometry and tetrahedral coordination framework. Because both polymorphs are noncentrosymmetric and direct-gap semiconductors \cite{laref1999tight}, the comparison is not limited to a symmetry-allowed versus symmetry-forbidden SHG case. Instead, it permits a more selective examination of how a change in long-range crystal symmetry modifies the angular response of both SHG and 2PPL within a common chemical system.

In this work, phase-dependent nonlinear optical response in AgI thin films synthesized by controlled physical vapor deposition is investigated using polarization-resolved SHG and 2PPL spectroscopy. Distinct angular responses are observed for triangular crystals assigned to \zbAgI (111) and rod-like crystals assigned to \wzAgI (101). We thus establish that in a single-composition binary halide, both symmetry and distinct morphological effects on $\chi^{(2)}$ (SHG) and $\chi^{(3)}$ (2PPL) can be analyzed simultaneously. The values of second-order nonlinear susceptibilities $\chi^{(2)}_\mathrm{eff}$ were estimated to be 0.45 pm/V (\zbAgI) and  0.16 pm/V (\wzAgI), respectively, by using quartz as a reference and correcting for an interference effect coming from the AgI film and the dielectric substrate underneath. The calibration approach for $\chi^{(2)}$ adopted here should be useful to evaluate other emerging highly-nonlinear layered materials or thin films supported by a dielectric substrate. 

\section{Experimental}
\begin{figure*}[!htb]
\includegraphics[width=16cm,clip]{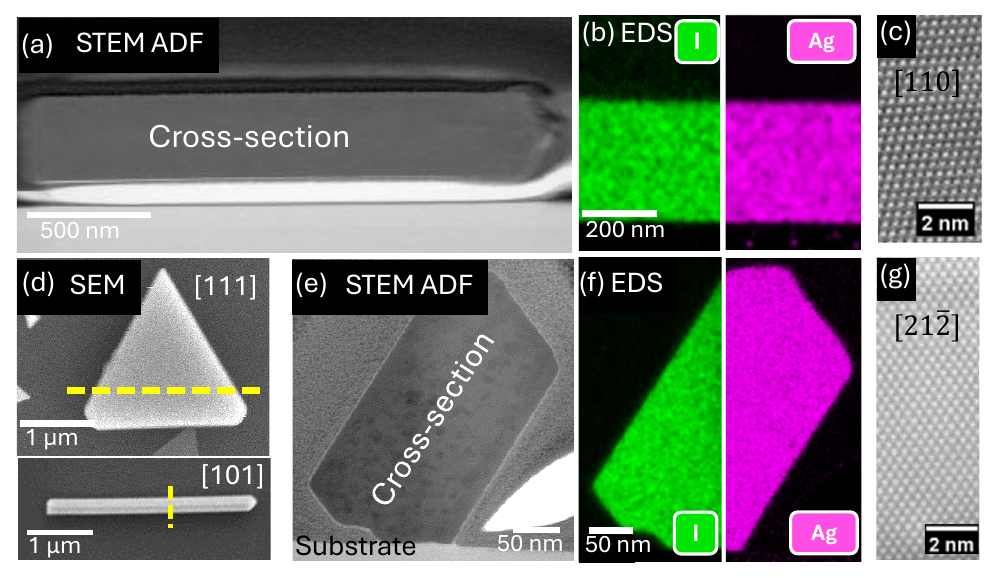}
\caption{%
Electron microscopy and EDS characterization of two AgI samples, a triangular platelet and a rod-like platelet, prepared as FIB lift-out cross-sectional lamellae. Triangular platelet: (a) STEM-ADF image of the cross section. (b) EDS elemental map showing Ag (magenta) and I (green). (c) High-resolution STEM image showing the [110] zone axis, consistent with $\gamma$-AgI. (d) SEM plan-view images of the triangular and rod-like platelets used for TEM sample preparation. The dashed yellow lines indicate the cutting locations for cross sectioning. Rod-like platelet: (e) STEM-ADF image of the cross section. (f) EDS elemental map. (g) High-resolution STEM image showing the $[21\bar{2}]$ zone axis, consistent with $\beta$-AgI.
}
\label{fig:morphology}
\end{figure*}

Silver iodide (AgI) microcrystals were synthesized in a custom-built quartz-tube physical vapor deposition system based on a horizontal furnace (MTI) with an 8-inch-long heating zone mounted on a sliding rail. Approximately 0.1 g of high-purity AgI powder was loaded on a ceramic plate, and three 1 cm $\times$ 1 cm Si/SiO$_2$ substrates with a 90 nm oxide layer were placed downstream on a separate ceramic plate. Before growth, the tube was evacuated for 5 min, purged with a 96\% Ar and 4\% H$_2$ gas mixture up to 0.02 MPa, evacuated again for 5 min, and then placed under carrier-gas flow. We adopted reverse flow (substrate to powder) during the temperature ramp to prevent growth, and switched to forward flow (powder to substrate) at the target growth temperatures, which were 450 $^\circ$C for triangular platelets, and 470 $^\circ$C for rod-shaped micro-crystals, respectively. Growth was terminated by returning to reverse flow, rapidly sliding the furnace away from the substrate region, and then cooling to room temperature under carrier gas.

Cross-sectional lamellae from representative triangular and rod-like crystals were prepared by focused-ion-beam lift-out using an FEI Scios DualBeam FIB/SEM, followed by final polishing at 2 keV to reduce surface damage. Scanning transmission electron microscopy was carried out on a probe-corrected Titan Themis 80–300 operated at 300 kV with a probe semi-convergence angle of 23.8 mrad and a beam current of 80 pA. Annular dark-field and high-angle annular dark-field imaging, together with Bruker SuperX energy-dispersive X-ray spectroscopy mapping, were used to evaluate morphology and elemental distribution. Phase assignment was based on lattice-resolved STEM, fast Fourier transforms, and zone-axis indexing, with comparison to expected interplanar spacings for candidate AgI polymorphs and, where applicable, simulated patterns. The triangular platelets were found to be consistent with $\gamma$-AgI with a nominal (111) exposed surface, whereas the rod-like crystals were found to be consistent with $\beta$-AgI with a nominal (101) exposed surface. For the rod-like crystals, the TEM indexing was corroborated by XRD, which showed a peak at $2\theta \approx 25.3^\circ$ consistent with $\beta$-AgI(101). Representative triangular platelets exhibited an approximately 500 nm cross-sectional thickness.
 
Polarization-resolved second-harmonic generation (SHG) and two-photon photoluminescence (2PPL) measurements were performed in a custom reflection-geometry setup described in Ref.\cite{puri2024substrate} using a mode-locked Ti:sapphire laser (Tsunami, Spectra-Physics).  An ultrafast pulsed laser with a pulse width of 80 fs and a near-infrared wavelength of $780 - 830$ nm was used as an incident beam. A prism-based compressor was used to maintain comparable pulse conditions over the tuning range. A dichroic mirror directed the excitation beam through a 40$\times$ objective (NA 0.6), which gave a spot diameter of approximately 1.6 $\mu$m at $\lambda \approx 800$ nm. The excitation power was set to 1.5 mW for polarization-dependent measurements. White-light illumination, a beam splitter, a tube lens, and a CCD camera were used for sample identification and alignment. The emitted nonlinear signal was collected through the same objective, routed through spectral filters to suppress the fundamental beam, and recorded with a spectrometer (Acton Research) equipped with a thermoelectrically cooled CCD (Horiba). The spectrometer with CCD was calibrated with a mercury pen lamp.

For angular measurements, the incident polarization was fixed by a linear polarizer in the excitation path, the sample was rotated with respect to a fixed pump polarization, and a second linear polarizer in the detection path selected the emitted component parallel to the incident polarization. The intensity of SHG and 2PPL was analyzed by integrating each emission over a corresponding spectral window. Rod-like crystals occasionally exhibited time-dependent signal degradation during rotation scans. In those cases, the analysis was restricted to the stable portion of the scan, and the angular traces were normalized to their own maxima before fitting.

\section{Results and discussion}\label{res}

\subsection*{Phase-selective SHG and excitonic 2PPL}
\begin{figure}[!htb]
\includegraphics[width=8.5cm,clip]{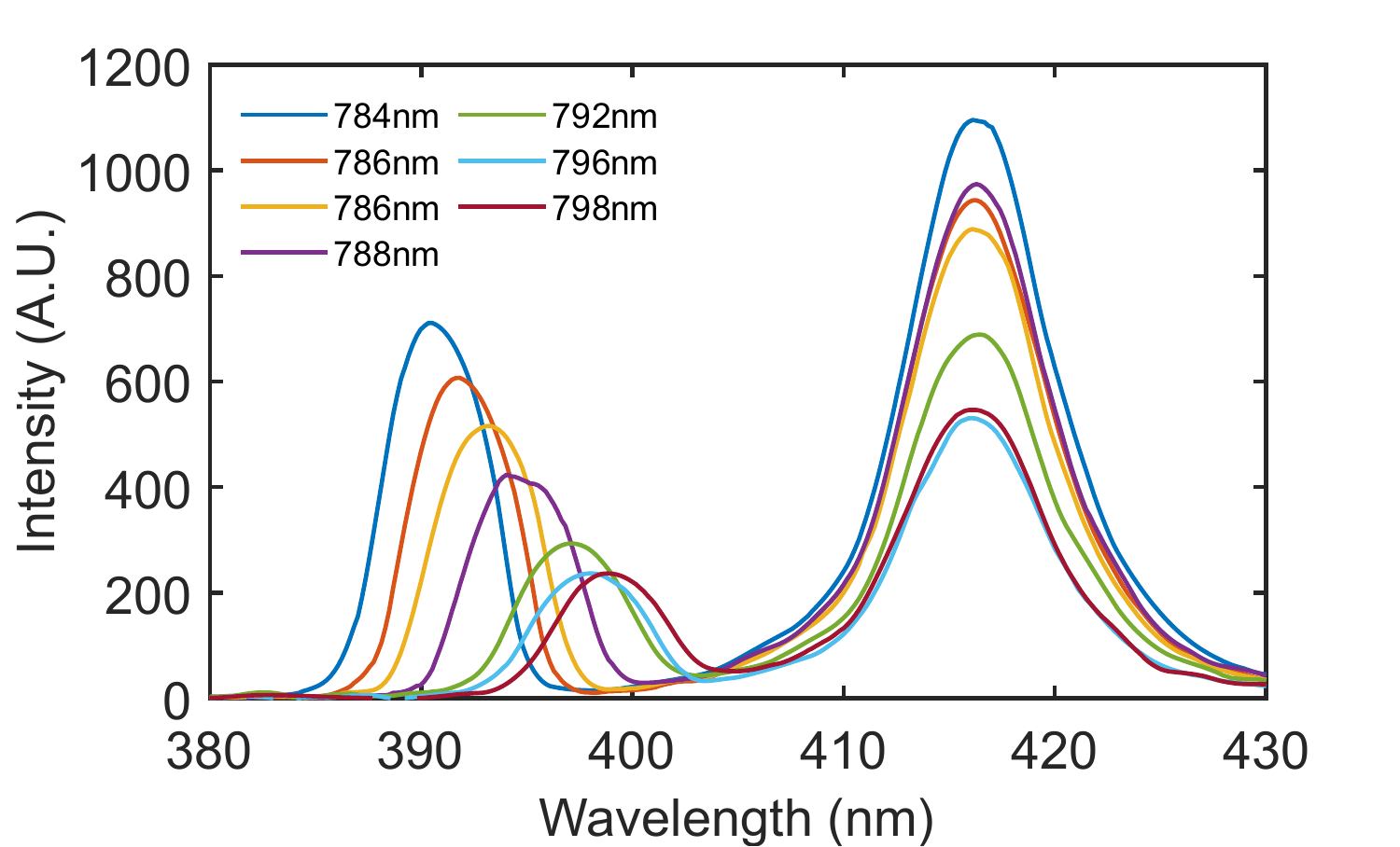}
\caption{%
Wavelength-resolved nonlinear emission from an individual AgI triangular flake. Photoluminescence spectra were collected as the excitation wavelength was tuned from 784 nm to 798 nm. Each trace exhibits two distinct features: (i) a band at 388–398 nm attributed to second-harmonic generation (SHG), which shifts according to $\lambda/2$ of the fundamental excitation; and (ii) another band at 412–420 nm assigned to excitonic emission, whose peak position remains nearly constant while its intensity varies with the pump wavelength. All intensities are plotted in arbitrary units (A.U.).
}
\label{fig:rawData}
\end{figure}

Figure~\ref{fig:rawData} presents wavelength-resolved nonlinear emission spectra from an individual AgI triangular flake. The spectra reveal two distinct peaks: (i) a shorter-wavelength band at 388–398 nm corresponding to SHG, confirmed by its $\lambda_{out} = \lambda_{in}/2$ dependence, and (ii) another peak at approximately 416 nm, assigned to the excitonic transition. The exciton emission varied from 412 nm to 440 nm, depending on the specific flake measured, likely originating from different crystal phases as detailed below. Since the fundamental laser energy is well below the estimated bandgap of around 3 eV in AgI \cite{goldmann1977band}, the observed exciton peak is attributed to 2PPL. 

\subsection*{Reference-based estimate of $\chi^{(2)}$}

\begin{figure}[!htb] 
\includegraphics[width=8.5cm,clip]{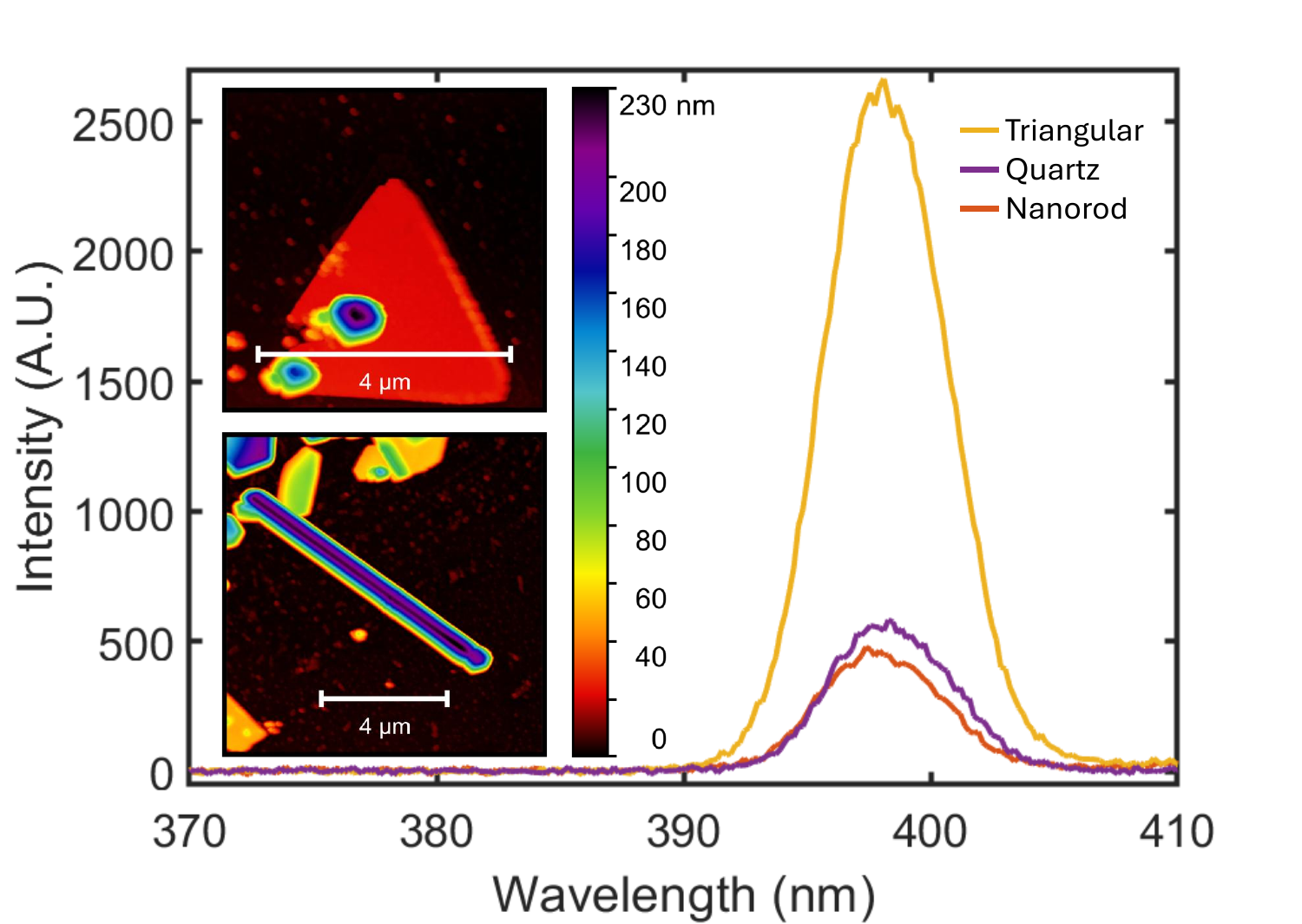} 
\caption{SHG spectra measured from representative AgI crystals with triangular and nanorod morphologies, together with a commercial quartz reference [Quartz (0001), 5 mm $\times$ 5 mm $\times$ 0.5 mm]. Insets show AFM topography images of the corresponding triangular platelet and nanorod flakes.} 
\label{fig:chi2} 
\end{figure}

A quantitative estimate of the second-order nonlinear susceptibility was obtained by comparing the SHG intensity of AgI crystals with that of a commercial $\alpha$-quartz reference, measured sequentially under the same nominal excitation, integration time, spectrometer settings, and polarization configuration. The incident fundamental power was 1.0~mW for AgI and 2.0~mW for quartz, which was accounted for using the quadratic power dependence of SHG. Minor refocusing and sample-angle adjustments were used to maximize the collected signal from each specimen (Fig.~\ref{fig:chi2}). 

Because the AgI crystals possess thicknesses comparable to the optical wavelengths inside the material (e.g., 220~nm for the nanorods), an infinitesimally thin sheet approximation is insufficient. Following Lavrov \textit{et al.}, we employ a depth-integrated formalism \cite{Lavrov2016SHG} to account for the phase-mismatch and multiple internal reflections of the fundamental and second-harmonic fields within the bulk AgI/SiO$_2$/Si stack. The absolute value of $\chi^{(2)}_{\mathrm{AgI}}$ is extracted by comparing the integrated AgI response to a bulk $\alpha$-quartz reference, utilizing a model that incorporates the focused Gaussian geometry of the beam to appropriately handle the finite coherent buildup within the focal volume \cite{Malard2013observation}.

By taking the ratio of the generated SHG efficiencies, as described in detail in the Appendix E, the comparative expression for the effective bulk susceptibility becomes:
\begin{equation}
\begin{split}
|\chi^{(2)}_{\mathrm{AgI}}| ={}& |\chi^{(2)}_{q}| \left[ \frac{16 \lambda^2 f^2 (\mathrm{N.A.})^4 n_{\mathrm{AgI}}(2\omega)^2}{d_L^2 G_{\mathrm{MR}} n_q^2 (n_q+1)^6} \right. \\
&\times \left. \frac{I_{\mathrm{AgI}}(2\omega)}{I_q(2\omega)} \left(\frac{P_q}{P_{\mathrm{AgI}}}\right)^2 \right]^{1/2},
\end{split}
\label{eq:chi2_ratio}
\end{equation}
where $f=0.056$ is the numerical integration constant for the focused Gaussian beam \cite{Malard2013observation}, $\mathrm{N.A.}$ is the objective numerical aperture, $\lambda$ is the fundamental wavelength in vacuum, and $d_L$ is the thickness of the AgI layer. The term $G_{\mathrm{MR}}$ is a dimensionless structure factor that integrates the depth-dependent local pump field enhancement and second-harmonic out-coupling influenced by multiple internal reflections within the target stack. $I_{\mathrm{AgI}}(2\omega)$ and $I_q(2\omega)$ denote the maximum SHG peak counts, and $P_{\mathrm{AgI}}$ and $P_q$ are the respective incident pump powers. The detailed derivation of the integration and structure factor is deferred to Appendix~\ref{app:chi2_model}.

Using this approach, the extracted bulk susceptibilities were estimated as:
\begin{equation}
\chi^{(2)}_{\mathrm{AgI(zb)}} \approx 0.45~\mathrm{pm/V},
\end{equation}
for the 19~nm zincblende flake, and
\begin{equation}
\chi^{(2)}_{\mathrm{AgI(wz)}} \approx 0.16~\mathrm{pm/V},
\end{equation}
for the 220~nm wurtzite rod. These bulk values fall well within the typical range for wide-gap nonlinear crystals such as KDP ($\chi^{(2)}\sim0.42~\mathrm{pm/V}$) and quartz ($\chi^{(2)}\sim0.6~\mathrm{pm/V}$) \cite{Boyd2020Book}. We note that the same refractive-index values were used for zincblende and wurtzite AgI because phase-resolved optical constants for zincblende AgI are unavailable. Thus, the quoted values should be regarded as approximate comparative estimates.

\subsection*{Analysis of SHG and 2PPL Angular Dependence}

\begin{figure*}[!htb]
\includegraphics[width=14cm,clip]{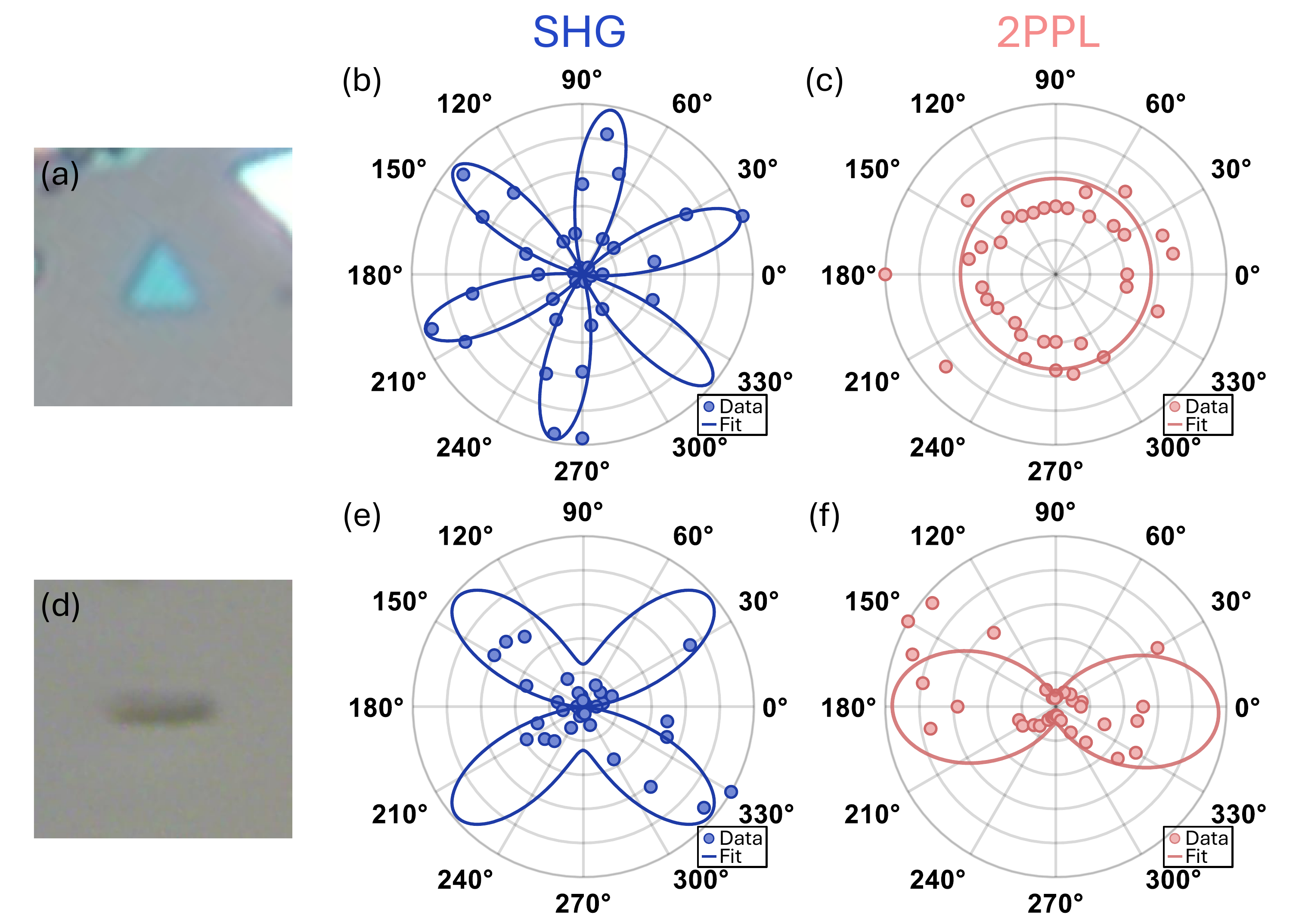}
\caption{%
Polarization-resolved nonlinear-optical response of two AgI micro-crystals.
(a) Optical micrograph of a triangular flake; (b) SHG polar plot from flake (a), exhibiting a six-lobed pattern that follows a $\cos^2[3(\theta-\phi_0)]$ dependence; (c) 2PPL response of the triangle. (d) Optical micrograph of a nanorod; (e) SHG polar plot recorded from the nanorod showing twofold symmetry aligning with simulation values for ${d_{1}>d_{2}}$; (f) corresponding 2PPL revealing pronounced anisotropy, contrasting the isotropic response from (a) – maxima are aligned with the long axis of the rod.
}
\label{fig:Polar}
\end{figure*}

To study phase-dependent nonlinear anisotropy in AgI, we have performed polarization-resolved SHG and 2PPL measurements for samples with different morphologies: triangular platelets and rod-shaped AgI.
Distinct angular anisotropy was observed in SHG from those crystals, exhibiting a sixfold ``flower'' pattern for the triangular AgI (Fig.\ref{fig:Polar}(b)) compared to a twofold ``butterfly'' pattern for the AgI rod (Fig.\ref{fig:Polar}(e)). The 2PPL also showed distinct behavior depending on the morphologies: We observe an isotropic circular response for the triangular island (Fig.\ref{fig:Polar}(c)) and a twofold uniaxial pattern in the rod (Fig.\ref{fig:Polar}(f)).  Note that although triangular AgI consistently displayed uniform angular-SHG patterns, the patterns derived from rod-shaped AgI were more diverse, sometimes also indicating the impact of beam damage. 

To analyze the nonlinear anisotropy observed in zincblende and wurtzite phases of AgI with distinct morphologies, we employ second- and third-order nonlinear optical formalisms that account for both crystal symmetry and the experimental geometry.

\begin{figure}[!htb]
\includegraphics[width=\linewidth,clip]{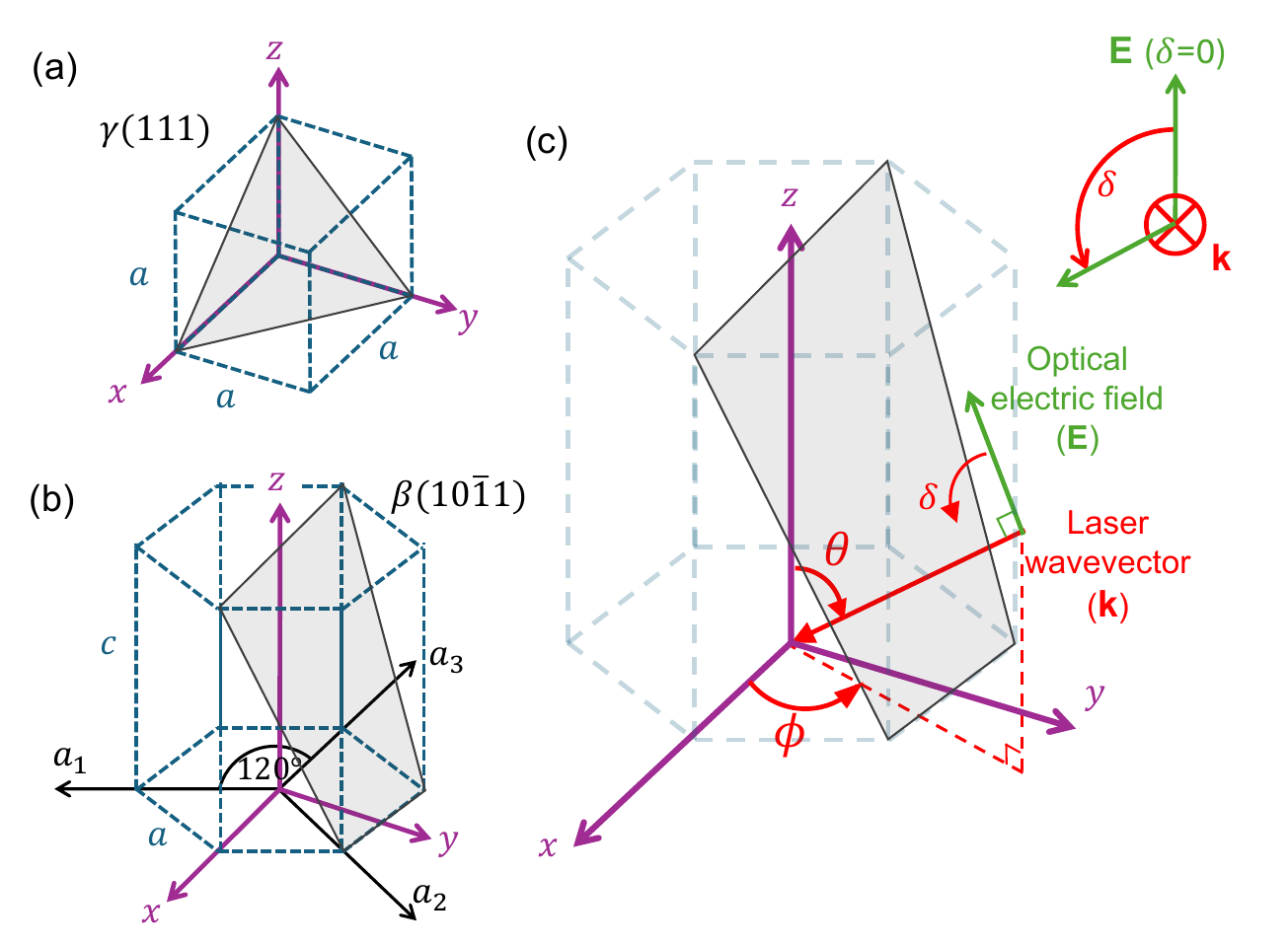}
\caption{%
Relation between Cartesian and crystallographic coordinate systems, and definition of beam orientation angles.
(a) Alignment of Cartesian and crystallographic axes in a cubic unit cell. 
(b) Alignment of Cartesian and crystallographic axes in a hexagonal unit cell. 
(c) Definition of the angles $\theta$, $\phi$, and $\delta$ describing the orientation of the incident laser beam (wave vector $\mathbf{k}$) and its optical electric field $\mathbf{E}$.
}
\label{fig:coordinates}
\end{figure}

\paragraph*{Second-harmonic generation (SHG).}
The three orthogonal components of the second-order nonlinear polarization $P_{i,2\omega}$ are expressed as
\begin{equation}\label{eq:Pol}
P_{i,2\omega} = 2\varepsilon_0 \sum_{j,k} d_{ijk}^{(2)}(2\omega;\omega,\omega)\, E_j E_k,
\end{equation}
where $\varepsilon_0$ is the permittivity of free space, 
$d_{ijk}^{(2)}=\tfrac{1}{2}\chi_{ijk}^{(2)}$ is the second-order susceptibility tensor, 
and $E_jE_k$ are components of the fundamental electric field.

For the zincblende phase of AgI (point group 4$\overline{3}$m), 
the nonvanishing elements of $d_{ijk}$ are~\cite{Boyd2020Book}
$d_{xyz}=d_{yzx}=d_{zxy}\equiv d_{14}$, assuming $d_{ijk}=d_{ikj}$ symmetry: 
\begin{equation}\label{eq:zbTensor}
\mathbf{d}_{\mathrm{zb}} =
\begin{bmatrix}
0 & 0 & 0 & d_{14} & 0 & 0\\
0 & 0 & 0 & 0 & d_{14} & 0\\
0 & 0 & 0 & 0 & 0 & d_{14}
\end{bmatrix}.
\end{equation}
For the wurtzite phase (point group 6mm), the nonvanishing elements are 
$d_{zxx}=d_{zyy}=d_{yyz}=d_{xxz}\equiv d_{15}$ 
and $d_{zzz}\equiv d_{33}$~\cite{Boyd2020Book}:
\begin{equation}\label{eq:wzTensor}
\mathbf{d}_{\mathrm{wz}} =
\begin{bmatrix}
0 & 0 & 0 & 0 & d_{15} & 0\\
0 & 0 & 0 & d_{15} & 0 & 0\\
d_{15} & d_{15} & d_{33} & 0 & 0 & 0
\end{bmatrix}.
\end{equation}
The $x$, $y$, and $z$ labels correspond to laboratory axes aligned with the crystallographic axes 
as defined in Fig.~\ref{fig:coordinates}.

To analyze polarization-resolved SHG when $\mathbf{E}$ is rotated within a crystallographic plane, 
we introduce an effective scalar susceptibility
\begin{equation}\label{eq:deff}
d_{\mathrm{eff}}(\theta,\phi,\delta) =
(e_{3x},e_{3y},e_{3z})\cdot
\mathbf{d}\!
\begin{bmatrix}
e_{1x}^2\\ e_{1y}^2\\ e_{1z}^2\\ 2e_{1y}e_{1z}\\ 2e_{1x}e_{1z}\\ 2e_{1x}e_{1y}
\end{bmatrix},
\end{equation}
where $\bm{e}_3$ denotes the SH polarization direction after the analyzer, 
and $\bm{e}_1$ is the polarization of the fundamental beam. 
In our configuration $\bm{e}_3=\bm{e}_1=\bm{e}$, described by
\begin{equation}\label{eq:evec}
\begin{split}
e_x &= \sin\phi\sin\delta - \cos\theta\cos\phi\cos\delta,\\
e_y &= -\cos\phi\sin\delta - \cos\theta\sin\phi\cos\delta,\\
e_z &= \sin\theta\cos\delta.
\end{split}
\end{equation}
Rotation of polarization within the zincblende $(111)$ plane corresponds to rotating $\delta$ from 0$^\circ$ to 360$^\circ$
at fixed $\theta=54.74^\circ$, $\phi=45^\circ$, 
while for wurtzite $(101)$ $\theta=62^\circ$ and $\phi=30^\circ$. 
The scalar quantity $d_{\mathrm{eff}}(\theta,\phi,\delta)$ can thus be computed and scaled to fit the SHG angular pattern.

\paragraph*{Two-photon absorption (2PA) and two-photon photoluminescence (2PPL).}
The 2PA process \cite{sheik1990sensitive,DeSalvo1993} originates from the imaginary part of the third-order nonlinear susceptibility, $\chi^{(3)}$, 
which governs the third-order polarization
\begin{equation}
P_i^{(3)} = \varepsilon_0 \sum_{jkl} \chi^{(3)}_{ijkl}\,E_jE_kE_l.
\end{equation}
The time-averaged absorbed power \cite{landau1984electrodynamics} is 
\begin{equation}
\label{eq:power_abs}
\langle \mathbf{E}\!\cdot\!\dot{\mathbf{P}}^{(3)}\rangle 
\propto \mathrm{Im}\!\left[\chi^{(3)}_{\mathrm{eff}}(\theta,\phi,\delta)\right]\,|E|^4,
\end{equation}
where we define the \emph{effective scalar susceptibility}
\begin{equation}
\label{eq:chieff_def}
\begin{split}
\chi^{(3)}_{\mathrm{eff}}(\theta,\phi,\delta)
&\equiv 
\sum_{ijkl}
\chi^{(3)}_{ijkl}\,
e_i(\theta,\phi,\delta)\,e_j(\theta,\phi,\delta)
\\[-4pt]&\hspace{2.3em}\times
e_k(\theta,\phi,\delta)\,e_l(\theta,\phi,\delta).
\end{split}
\end{equation}
This contraction reduces the fourth-rank tensor $\chi^{(3)}_{ijkl}$ 
to a complex scalar quantity that depends explicitly on the optical geometry $(\theta,\phi,\delta)$. 
This $E^4$ scaling is consistent with both quantum-mechanical calculations of 2PA matrix elements~\cite{Nascimento1983Polarization}
and macroscopic susceptibility analyses~\cite{Dvorak1994Measurement,Hutchings1994Polarization}.
The measured 2PPL intensity is assumed to be proportional to the absorbed two-photon energy and unpolarized due to inelastic scattering before emission, giving
\begin{equation}
I_{2\mathrm{PPL}}(\theta,\phi,\delta)\ \propto\ I_{2\mathrm{PA}}(\theta,\phi,\delta).
\end{equation}

\medskip
\noindent
Under degenerate excitation with linear polarization, the relevant quantity in Eq.~(\ref{eq:chieff_def}) is the quartic field product $E_iE_jE_kE_l$.  
Because the two absorbed photons are indistinguishable, the indices associated with the input fields $(j,k,l)$ enter symmetrically.
We therefore group the quadratic field products into a six-component vector,
\begin{equation}
v(\delta)=
\begin{bmatrix}
E_x^2 & E_y^2 & E_z^2 & 2E_yE_z & 2E_zE_x & 2E_xE_y
\end{bmatrix}^{\mathsf T},
\end{equation}
so that the 2PA signal may be written compactly as a quadratic form
\begin{equation}
\label{eq:I2PA_voigt}
I_{2\mathrm{PA}}(\theta,\phi,\delta)\ \propto\ v(\delta)^{\mathsf T} B\,v(\delta).
\end{equation}
Here, $B$ is a real, symmetric $6\times6$ matrix whose entries are linear combinations of the independent components of $\mathrm{Im}[\chi^{(3)}_{ijkl}]$ \cite{Hutchings1994Polarization, Karkhanehchi1997Polarization}. The matrix $B$ encapsulates all crystal-symmetry constraints and fully determines the angular dependence of 2PA through the direction cosines $(E_x, E_y, E_z)=(e_x,e_y,e_z)$ in Eq.~(\ref{eq:evec}).
This is the same structure that underlies Voigt contraction for other fourth-rank optical tensors (e.g., Kerr and elastic tensors \cite{nye1985physical,yariv1984optical})). Although not frequently used in the context of 2PA, it makes crystal-symmetry constraints (cubic vs.\ hexagonal) transparent. Angular anisotropy of 2PA under more general conditions is discussed, for example, in Refs. \cite{Dvorak1994Measurement,Hutchings1994Polarization,KraussKodytek2021ZnSe}.
  
Expanding Eq.~(\ref{eq:I2PA_voigt}) gives
\begin{multline}
\label{eq:I2PA_expand}
I_{2\mathrm{PA}}\propto
B_{11}E_x^4 + B_{22}E_y^4 + B_{33}E_z^4 \\[2pt]
+\,2\!\left(B_{12}E_x^2E_y^2 + B_{13}E_x^2E_z^2 + B_{23}E_y^2E_z^2\right) \\[2pt]
+\,B_{44}(2E_yE_z)^2 + B_{55}(2E_zE_x)^2 + B_{66}(2E_xE_y)^2.
\end{multline}

Following the tensor analysis of Inoue and Toyozawa \cite{inoue1965two} for two-photon absorption in semiconductors, the fourth-rank susceptibility entering the 2PPL process can be decomposed into a small number of scalar invariants fixed by crystal symmetry. For zincblende, the two-photon absorption tensor can be written as a sum of only two linearly independent invariants: one isotropic term and one cubic-anisotropy term. 
\begin{equation}
  \beta_1^{(\mathrm{zb})} = B_{11} + 2 B_{12}, 
  \qquad
  \beta_2^{(\mathrm{zb})} = B_{11} - \bigl(B_{11} + 2 B_{12}\bigr) \, 
\end{equation}
Here, $\beta^{(\mathrm{zb})}_1$ sets the isotropic background, while $\beta^{(\mathrm{zb})}_2$ controls the cubic anisotropy.
In contrast, the lower hexagonal symmetry of wurtzite allows four independent invariants in the same decomposition. The corresponding 2PPL intensity then depends on four independent scalar coefficients, each a distinct linear combination of the underlying tensor elements.
\begin{equation}
\begin{split}
\beta_1^{(\mathrm{wz})} &= B_{33}, \qquad
\beta_2^{(\mathrm{wz})} = 2 B_{13} + 4 B_{44},\\
\beta_3^{(\mathrm{wz})} &= B_{11}, \qquad
\beta_4^{(\mathrm{wz})} = 2 B_{12} + 4 B_{66} - 6 B_{11}
\end{split}
\end{equation}
Here, $\beta^{(\mathrm{wz})}_3$ governs the in-plane isotropic part and $\beta^{(\mathrm{wz})}_4$ the in-plane uniaxial (twofold) anisotropy, whereas $\beta^{(\mathrm{wz})}_1$ and $\beta^{(\mathrm{wz})}_2$ weight the out-of-plane and mixed ($E_z$) contributions.

\paragraph*{Zincblende (4$\overline{3}$m) form of $B$.}
For cubic (zincblende) symmetry, the tensor simplifies to
\begin{equation}
\label{eq:B_43m}
B_{zb}=
\begin{bmatrix}
B_{11} & B_{12} & B_{12} & 0 & 0 & 0\\
B_{12} & B_{11} & B_{12} & 0 & 0 & 0\\
B_{12} & B_{12} & B_{11} & 0 & 0 & 0\\
0 & 0 & 0 & B_{44} & 0 & 0\\
0 & 0 & 0 & 0 & B_{44} & 0\\
0 & 0 & 0 & 0 & 0 & B_{44} 
\end{bmatrix},
\end{equation}
a form commonly used for cubic semiconductors exhibiting zincblende-type symmetry~\cite{Hutchings1994Polarization, KraussKodytek2021ZnSe}.

\paragraph*{Wurtzite (6mm) form of $B$.}
For crystals with 6mm symmetry, shear planes containing the $c$-axis are equivalent, giving
\begin{equation}
\label{eq:B_6mm}
B_{wz}=
\begin{bmatrix}
B_{11} & B_{12} & B_{13} & 0 & 0 & 0\\
B_{12} & B_{11} & B_{13} & 0 & 0 & 0\\
B_{13} & B_{13} & B_{33} & 0 & 0 & 0\\
0 & 0 & 0 & B_{44} & 0 & 0\\
0 & 0 & 0 & 0 & B_{44} & 0\\
0 & 0 & 0 & 0 & 0 & B_{66}
\end{bmatrix}.
\end{equation}

The results of theoretical fits to SHG and 2PPL using the models described above are shown as solid lines in Fig.~4. For zincblende (111), the SHG flower pattern is reproduced well, and the nearly circular 2PPL response is reproduced with $|\beta^{(\mathrm{zb})}_{2}| \ll |\beta^{(\mathrm{zb})}_{1}|$, consistent with earlier results on GaAs(111) \cite{Dvorak1994Measurement}. Fits to the rod data, on the other hand, require $|\beta^{(\mathrm{wz})}_4|\gg|\beta^{(\mathrm{wz})}_3|$ with a finite $\beta^{(\mathrm{wz})}_2$, consistent with the observed twofold pattern and its axis.

For the rod-like crystals, the agreement between model and experiment is qualitative rather than exact. The calculated pattern reproduces the dominant twofold character of the SHG response, but residual deviations remain in lobe amplitude and angular position (see Fig.~4 and Appendix Fig.~7). Several factors may contribute to this mismatch. The analytical treatment of angular response assumes a uniform bulk crystal.  It does not explicitly account for finite rod geometry, edge effects, or morphology-induced redistribution of the optical near field. For example, the electric field components polarized perpendicular to the long axis of the rod could be diminished due to geometrical effects \cite{bautista2015second,wiecha2016origin,neeman2017crystallographic,johnston2022polarization}. In addition, the rod measurements were more susceptible to degradation during acquisition, and small changes in alignment or sample condition can distort the angular trace. The theoretical fits for Wurtzite should therefore be interpreted as capturing the leading symmetry-derived anisotropy and the exact ratio of different tensor elements in $\chi^{(2)}$ or $\chi^{(3)}$ (through $\beta$ terms) can be quantitatively different from a uniform bulk case.

Although the optical trends correlate well with the structural assignments, the nonlinear anisotropy should not be interpreted as a sole phase fingerprint. The phase assignment is supported primarily by lattice-resolved STEM, zone-axis analysis, and, for the rod-like crystals, corroborated by XRD. Although the present data suggests that the crystallographic phase is a dominant factor in determining SHG and 2PPL anisotropy, morphology-related effects, particularly in the rod-like crystals, may contribute to the exact amplitude of lobes forming the measured anisotropy.

\section{Summary}
In conclusion, we have investigated the nonlinear optical properties of silver iodide (AgI) thin films using polarization-resolved second harmonic generation and two-photon photoluminescence spectroscopy. By tuning the physical vapor deposition (PVD) conditions, we selectively grew zincblende (\zbAgI) and wurtzite (\wzAgI) phases. Our measurements reveal that triangular \zbAgI $(111)$ flakes exhibit sixfold SHG and isotropic 2PPL angular patterns, while rod-like \wzAgI $(101)$ crystals display a butterfly-shaped SHG and twofold 2PPL angular patterns. These contrasting behaviors are consistent with theoretical analysis and underscore the role of crystal phase in dictating nonlinear optical behavior. The ability to engineer the phase and resultant morphology of AgI provides a promising platform for tailoring optical anisotropy and nonlinear light–matter interactions at the nanoscale.

\begin{acknowledgments}
We thank N. Nagaosa for helpful discussions. The work is supported by the Office of the Secretary of Defense for Research and Engineering under Award No. FA9550-23-1-0500 and the MonArk NSF Quantum Foundry from the National Science Foundation under Award No. DMR-1906383. L. L. S. was supported by the University of Arkansas Honors College Research Grant and the Chan and Chen Endowed Research Scholarship.
\end{acknowledgments}


\appendix

\section{Transmission Electron Microscopy Analysis}
\label{app:tem}

\subsection{Sample Preparation and Acquisition Conditions}

Cross-sectional lamellae from a representative triangular platelet and a representative rod-like crystal were prepared using an FEI Scios DualBeam FIB/SEM (Thermo Fisher Scientific). Final polishing at 2 keV was used to minimize surface damage. STEM analysis was performed at 300 kV on a probe-corrected Titan Themis 80--300 (Thermo Fisher Scientific) with a probe semi-convergence angle of 23.8 mrad and a beam current of 80 pA to reduce beam damage. Elemental maps were acquired using a Bruker SuperX EDS system.

\begin{figure*}[!htb]
\centering
\includegraphics[width=14cm,clip]{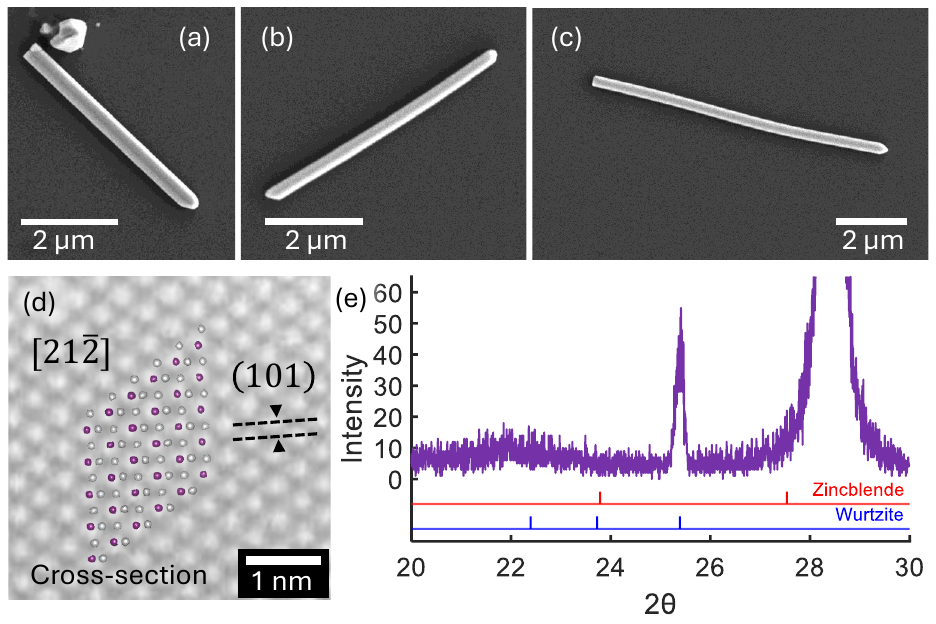}
\caption{%
Structural characterization of AgI nanorods (\wzAgI). (a)--(c) SEM images of representative AgI rods grown at $(470^\circ\mathrm{C}$, showing the elongated rod-like morphology. (d) HRSTEM image of a rod cross section acquired along the $[21\bar{2}]$ zone axis, with lattice planes indexed to the $(101)$ family. (e) X-ray diffraction (XRD) pattern of the rod-like sample. The peak at $2\theta = 25.3^\circ$ is consistent with the \wzAgI$(101)$ reflection and agrees with the TEM-based indexing, indicating a preferred $(101)$ orientation. The stronger peak near $2\theta = 28.5^\circ$ is assigned to the Si $(111)$ substrate.
}
\label{fig:app_S1}
\end{figure*}

\subsection{Lattice Indexing and Phase Assignment}

Figure~\ref{fig:app_S1}(d) shows a crystallographic model overlaid on the HRSTEM cross section acquired along the \([2\,1\,\overline{2}]\) zone axis, with lattice planes indexed to \((101)\). The XRD trace in Fig.~\ref{fig:app_S1}(e) corroborates this assignment, which shows a peak at \(2\theta = 25.3^\circ\) consistent with \wzAgI\((101)\) and the TEM analysis.

\section{Additional Polarization-Resolved SHG and 2PPL Data}
\label{app:polarization}

\begin{figure*}[!htb]
\centering
\includegraphics[width=13cm,clip]{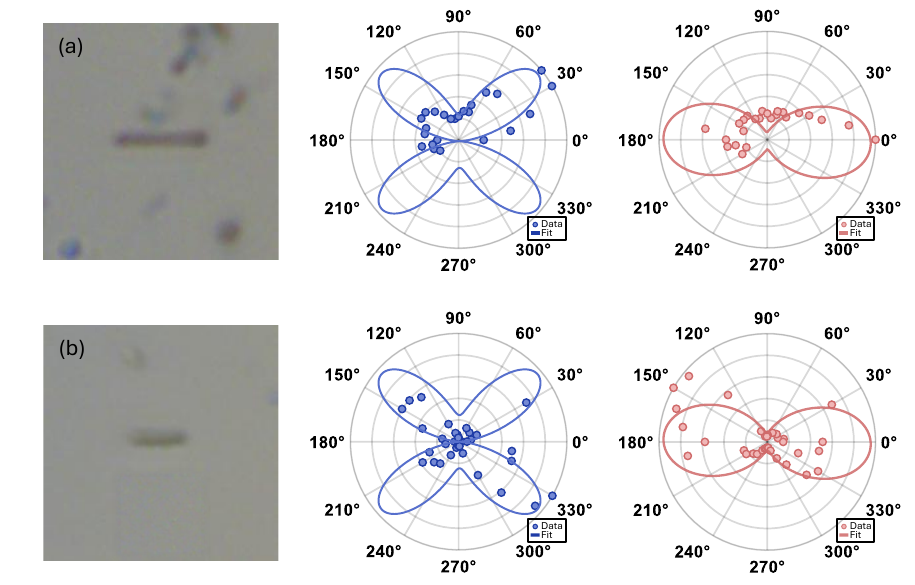}
\caption{%
Polarization-resolved SHG and 2PPL from AgI rods. For two representative rod-shaped crystals, polar plots of normalized SHG (blue dots) and 2PPL (pink dots) intensity are shown as a function of in-plane pump-polarization angle \(\delta\). SHG exhibits a butterfly-like twofold pattern, while 2PPL shows a uniaxial anisotropy with maximum intensity when the incident polarization is aligned with the long axis of the rod. All traces are normalized to their respective maxima.
}
\label{fig:app_S2}
\end{figure*}

Additional SHG and 2PPL data for AgI nanorods are shown in Fig.~\ref{fig:app_S2}. A predominantly twofold response was observed in thinner rods. The rod-shaped crystals frequently exhibited signs of degradation before \(180^\circ\), which limited the available angular coverage in some scans [Fig.~\ref{fig:app_S2}(a)]. The observed twofold pattern is consistent with simulations assuming a \wzAgI\((101)\) out-of-plane surface.

\section{Simulation Parameters for SHG}
\label{app:shg_sim}

\subsection{Zincblende}

\begin{equation}
\label{eq:app_zbTensor}
\mathbf{d}_{\mathrm{zb}} =
\begin{bmatrix}
0 & 0 & 0 & d_{14} & 0 & 0\\
0 & 0 & 0 & 0 & d_{14} & 0\\
0 & 0 & 0 & 0 & 0 & d_{14}
\end{bmatrix},
\qquad d_{14}=1.
\end{equation}

\subsection{Wurtzite}

\begin{equation}
\label{eq:app_wzTensor}
\begin{split}
\mathbf{d}_{\mathrm{wz}} &=
\begin{bmatrix}
0 & 0 & 0 & 0 & d_{15} & 0\\
0 & 0 & 0 & d_{15} & 0 & 0\\
d_{15} & d_{15} & d_{33} & 0 & 0 & 0
\end{bmatrix}, \\[1ex]
d_{15} &= 100, \quad d_{33} = 1.
\end{split}
\end{equation}

\section{Simulation Parameters for 2PPL}
\label{app:tppl_sim}

\subsection{Zincblende}

\begin{equation}
\label{eq:app_B43m}
\begin{bmatrix}
B_{11} & B_{12} & B_{12} & 0 & 0 & 0\\
B_{12} & B_{11} & B_{12} & 0 & 0 & 0\\
B_{12} & B_{12} & B_{11} & 0 & 0 & 0\\
0 & 0 & 0 & B_{44} & 0 & 0\\
0 & 0 & 0 & 0 & B_{44} & 0\\
0 & 0 & 0 & 0 & 0 & B_{44}
\end{bmatrix}.
\end{equation}

The parameter map used in the simulations was
\begin{equation}
\label{eq:app_B43m_map}
B_{11}=1, \qquad B_{12}=0, \qquad B_{44}=0.5,
\end{equation}
with
\begin{equation}
\beta_1^{(\mathrm{zb})}=1,
\qquad
\beta_2^{(\mathrm{zb})}=0.
\end{equation}

\subsection{Wurtzite}

\begin{equation}
\label{eq:app_B6mm}
B_{\mathrm{wz}}=
\begin{bmatrix}
B_{11} & B_{12} & B_{13} & 0 & 0 & 0\\
B_{12} & B_{11} & B_{13} & 0 & 0 & 0\\
B_{13} & B_{13} & B_{33} & 0 & 0 & 0\\
0 & 0 & 0 & B_{44} & 0 & 0\\
0 & 0 & 0 & 0 & B_{44} & 0\\
0 & 0 & 0 & 0 & 0 & B_{66}
\end{bmatrix}.
\end{equation}

The parameter map used in the simulations was
\begin{equation}
\label{eq:app_B6mm_map}
\begin{aligned}
B_{11} &= 0.5,   &\qquad B_{12} &= 0.2,   &\qquad B_{13} &= -0.2, \\
B_{33} &= 0.005, &\qquad B_{44} &= 0.12,  &\qquad B_{66} &= 0.1,
\end{aligned}
\end{equation}
with
\begin{equation}
\begin{aligned}
\beta_1^{(\mathrm{wz})} &= 0.005, & \beta_2^{(\mathrm{wz})} &= 0.08, \\[1ex]
\beta_3^{(\mathrm{wz})} &= 0.5, & \beta_4^{(\mathrm{wz})} &= -2.2.
\end{aligned}
\end{equation}

\section{Details of the Comparative Estimate of \texorpdfstring{$\chi^{(2)}$}{chi(2)}}
\label{app:chi2_model}

\begin{figure}[!htb]
\centering
\includegraphics[width=\columnwidth,clip]{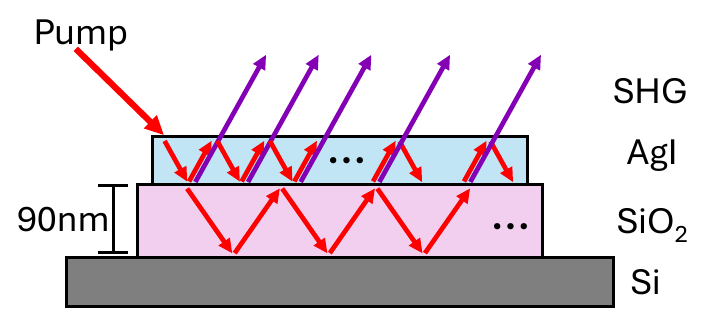}
\caption{%
Schematic of the multilayer optical geometry used to estimate the effective second-order susceptibility of AgI on \SiO/Si. The incident pump field undergoes multiple reflections within the AgI and SiO$_2$ layers, modifying the local fundamental field experienced by the AgI crystal. The generated second-harmonic field is emitted from the nonlinear AgI layer and is further affected by propagation and reflection in the optical stack.
}
\label{fig:app_S4}
\end{figure}

To accurately evaluate the second-order nonlinear susceptibility $\chi^{(2)}_{\mathrm{AgI}}$, we modeled the SHG intensity using a depth-integrated formalism that accounts for multiple internal reflections within the finite AgI layer \cite{Lavrov2016SHG}, followed by a direct comparison with a focused-beam calibration for bulk $\alpha$-quartz\cite{Malard2013observation}.

\subsection{Depth-Integrated Model for the AgI Layer}
We treat the sample as a multilayer optical stack (Fig.\ref{fig:app_S4}) consisting of air (0), the AgI layer ($L$) of thickness $d_L$, a SiO$_2$ oxide layer ($ox$) of thickness $d_{ox}$, and the Si substrate ($S$). The incident pump field (at frequency $\omega$) and the generated SHG field (at frequency $2\omega$) experience multiple internal reflections. Following an approach developed by Lavrov \textit{et al.} \cite{Lavrov2016SHG}, the reflected SHG field amplitude emitted back into the air ($x < 0$) is obtained by integrating the nonlinear polarization across the AgI thickness ($0 \leq x \leq d_L$):
\begin{equation}
\begin{split}
E_{2\omega}^{(\mathrm{ref})} ={}& \int_{0}^{d_{L}} \frac{i\omega}{n_{\mathrm{AgI}}(2\omega)c} \chi^{(2)}_{\mathrm{AgI}} E_{\mathrm{inc}}^{2} \\
&\times \left[u^{+}(x)+u^{-}(x)\right]^{2} F_{2\omega}(x) dx,
\end{split}
\label{eq:E2omega_ref}
\end{equation}
where $E_{\mathrm{inc}}$ is the incident pump electric field amplitude, $n_{\mathrm{AgI}}(2\omega)$ is the refractive index of AgI at the SHG frequency, and $c$ is the speed of light in vacuum. The dimensionless factors $u^+(x)$ and $u^-(x)$ describe the local enhancement of the forward and backward propagating fundamental fields at depth $x$:
\begin{align}
u^+(x) &= \frac{t_{0L}^{(1)} e^{ik_L^{(1)}x}}{1 - r_{L0}^{(1)} r_{L,ox,S}^{(1)} e^{2ik_L^{(1)}d_L}}, \\
u^-(x) &= \frac{t_{0L}^{(1)} r_{L,ox,S}^{(1)} e^{i k_L^{(1)} (2d_L - x)}}{1 - r_{L0}^{(1)} r_{L,ox,S}^{(1)} e^{2ik_L^{(1)}d_L}}.
\end{align}
Here, the superscript $(m=1,2)$ denotes the fundamental ($\omega$) and second-harmonic ($2\omega$) frequencies, respectively. The wavevectors are $k_j^{(m)} = \frac{2\pi m}{\lambda} n_j(m\omega)$, where $\lambda$ is the fundamental vacuum wavelength. The coefficients $t_{ij}^{(m)}$ and $r_{ij}^{(m)}$ are the standard Fresnel transmission and reflection coefficients at normal incidence between media $i$ and $j$. The effective reflection coefficient from the bottom interface structure (AgI/SiO$_2$/Si) is given by:
\begin{equation}
r_{L,ox,S}^{(m)} = \frac{r_{L,ox}^{(m)} + r_{ox,S}^{(m)} e^{2i k_{ox}^{(m)} d_{ox}}}{1 + r_{L,ox}^{(m)} r_{ox,S}^{(m)} e^{2i k_{ox}^{(m)} d_{ox}}}.
\end{equation}
Similarly, the out-coupling factor $F_{2\omega}(x)$, which dictates how the SHG field generated at depth $x$ propagates back to the air, is given by:
\begin{equation}
F_{2\omega}(x) = \frac{t_{L0}^{(2)} \left[ e^{i k_L^{(2)} x} + r_{L,ox,S}^{(2)} e^{i k_L^{(2)} (2 d_L - x)} \right]}{1 - r_{L0}^{(2)} r_{L,ox,S}^{(2)} e^{2i k_L^{(2)} d_L}}.
\end{equation}

By defining the SHG intensity in SI units as $I(2\omega) = \frac{1}{2}\epsilon_0 c |E_{2\omega}^{(\mathrm{ref})}|^2$ and isolating the fundamental intensity $I_{\mathrm{inc}}(\omega) = \frac{1}{2}\epsilon_0 c |E_{\mathrm{inc}}|^2$, the conversion efficiency for the AgI stack becomes:
\begin{equation}
\frac{I_{\mathrm{AgI}}(2\omega)}{I_{\mathrm{inc}}(\omega)^2} = \frac{2 \omega^2}{c^3 \epsilon_0 n_{\mathrm{AgI}}(2\omega)^2} d_L^2 G_{\mathrm{MR}} |\chi^{(2)}_{\mathrm{AgI}}|^2,
\end{equation}
where we have factored out the thickness to define the dimensionless optical structure factor $G_{\mathrm{MR}}$, which captures the total interference effect from the film and the dielectric layers beneath:
\begin{equation}
G_{\mathrm{MR}} = \left| \frac{1}{d_L} \int_{0}^{d_{L}} \left[u^{+}(x)+u^{-}(x)\right]^{2} F_{2\omega}(x) dx \right|^2.
\end{equation}

\subsection{Quartz Calibration and Final Formula}
For the bulk $\alpha$-quartz reference, the SHG intensity under a focused Gaussian beam geometry is described in SI units by\cite{Malard2013observation}:
\begin{equation}
\frac{I_{q}(2\omega)}{I_{\mathrm{inc}}(\omega)^2} = \frac{128\pi^2 f^2 (\mathrm{N.A.})^4}{c \epsilon_0 n_q^2 (n_q+1)^6} |\chi_q^{(2)}|^2,
\end{equation}
where $f=0.056$ is the numerical integration constant for a focused Gaussian beam, $\mathrm{N.A.}$ is the objective numerical aperture, and $n_q$ represents the quartz refractive indices. 

By taking the ratio of the AgI efficiency to the quartz efficiency, the fundamental constants ($\epsilon_0, c, \pi$) cancel. Accounting for the differences in experimental incident pump powers ($P_{\mathrm{AgI}}$ and $P_q$), the final analytical expression for the absolute susceptibility of AgI is:
\begin{equation}
\begin{split}
|\chi^{(2)}_{\mathrm{AgI}}| ={}& |\chi^{(2)}_{q}| \left[ \frac{16 \lambda^2 f^2 (\mathrm{N.A.})^4 n_{\mathrm{AgI}}(2\omega)^2}{d_L^2 G_{\mathrm{MR}} n_q^2 (n_q+1)^6} \right. \\
&\times \left. \frac{I_{\mathrm{AgI}}(2\omega)}{I_q(2\omega)} \left(\frac{P_q}{P_{\mathrm{AgI}}}\right)^2 \right]^{1/2}.
\end{split}
\label{eq:chi2_ratio_appendix}
\end{equation}

\begin{table}[htbp]
\centering
\caption{Constants and parameters used in the $\chi^{(2)}$ estimation. }
\label{tab:params}
\begin{tabular}{llrl}
\hline
Symbol & Quantity & Value & Unit \\
\hline
$\lambda$ & Fundamental wavelength & 800 & nm \\
$\chi_q^{(2)}$ & Quartz second-order susceptibility & 0.6 \cite{puri2024substrate} & pm/V \\
$n_q(\omega)$ & Quartz refractive index at $\omega$ & 1.4533 \cite{malitson1965interspecimen} & -- \\
$n_q(2\omega)$ & Quartz refractive index at $2\omega$ & 1.4701 \cite{malitson1965interspecimen} & -- \\
$n_{\mathrm{AgI}}(\omega)$ & AgI refractive index at $\omega$ & 2.045 \cite{ghosh1997handbook} & -- \\
$n_{\mathrm{AgI}}(2\omega)$ & AgI refractive index at $2\omega$ & 2.385 \cite{ghosh1997handbook} & -- \\
$f$ & Numerical integration constant & 0.056 & -- \\
$\mathrm{N.A.}$ & Objective numerical aperture & 0.6 & -- \\
$P_{\mathrm{AgI}}$ & AgI incident pump power & 1.0 & mW \\
$P_q$ & Quartz incident pump power & 2.0 & mW \\
$G_{\mathrm{MR(zb)}}$ & Zincblende structure factor & 4.626 & -- \\
$G_{\mathrm{MR(wz)}}$ & Wurtzite structure factor & 0.049 & -- \\
$I_{\mathrm{AgI(zb)}}(2\omega)$ & Zincblende SHG peak maximum & 2665 & counts \\
$I_{\mathrm{AgI(wz)}}(2\omega)$ & Wurtzite SHG peak maximum & 480 & counts \\
$I_q(2\omega)$ & Quartz SHG peak maximum & 584 & counts \\
$d_{L\mathrm{(zb)}}$ & Zincblende AgI thickness & 19 & nm \\
$d_{L\mathrm{(wz)}}$ & Wurtzite AgI thickness & 220 & nm \\
\hline
\end{tabular}
\vspace{0.8em}
\begin{minipage}{0.95\linewidth}
\footnotesize
\textit{Note.} The value of $n_{\mathrm{AgI}}(\omega)$ at the fundamental wavelength was obtained by extrapolation of the form
\[
n(\lambda)=1.93133808+\frac{0.05310596}{\lambda^{2}}+\frac{0.01243262}{\lambda^{4}},
\]
where $\lambda$ is expressed in $\mu\mathrm{m}$. 
\end{minipage}
\end{table}

\subsection{Numerical Evaluation}
To compute $\chi^{(2)}_{\mathrm{AgI}}$, we evaluated $G_{\mathrm{MR}}$ using the refractive indices, wavelengths, and thicknesses listed in Table~\ref{tab:params}. The numerical integration of the internal fields yields $G_{\mathrm{MR(zb)}} = 4.626$ for the $d_L = 19$~nm zincblende flake, and $G_{\mathrm{MR(wz)}} = 0.049$ for the $d_L = 220$~nm wurtzite nanorod. The vast reduction in $G_{\mathrm{MR(wz)}}$ reflects the severe internal phase-mismatch and Fabry-Perot interference suppressing the out-coupled SHG at that specific thickness.

Using the experimental raw counts ($I_{\mathrm{AgI(zb)}} = 2665$, $I_{\mathrm{AgI(wz)}} = 480$, $I_q = 584$) and applying the pump power scaling $(P_q/P_{\mathrm{AgI}})^2 = 4$, Eq.~\eqref{eq:chi2_ratio_appendix} directly yields:
\begin{equation}
\chi^{(2)}_{\mathrm{AgI(zb)}} \approx 0.45~\mathrm{pm/V},
\end{equation}
\begin{equation}
\chi^{(2)}_{\mathrm{AgI(wz)}} \approx 0.16~\mathrm{pm/V}.
\end{equation}
These values reflect the intrinsic bulk-like nonlinear response, separated from the geometric optical artifacts caused by the optical interference effects.


%

\end{document}